\documentclass{aa}
\usepackage{times}
\usepackage{graphics}
\usepackage{xspace}
\usepackage{epsfig}
\usepackage{jwaabib}
\usepackage{rotating}
\usepackage{dcolumn}

\begin{document}

\title{X-ray emission from a metal depleted accretion shock onto
the classical T Tauri star TW\,Hya}

\author{B. Stelzer\inst {1,2} \and J. H. M. M. Schmitt\inst {3}}

\institute{INAF - Osservatorio Astronomico di Palermo,
  Piazza del Parlamento 1,
  I-90134 Palermo,
  Italy \and
  Max-Planck-Institut f\"ur extraterrestrische Physik,
  Postfach 1312,
  D-85741 Garching,
  Germany \and
  Hamburger Sternwarte,
  Gojenbergsweg 112,
  D-21029 Hamburg, Germany}

\offprints{B. Stelzer}
\mail{B. Stelzer, stelzer@astropa.unipa.it}
\titlerunning{X-ray emission from an accretion shock on TW\,Hya}

\date{Received $<$date$>$ / Accepted $<$date$>$}

\abstract{ 
We present the X-ray spectrum of TW\,Hya observed at high and intermediate spectral
resolution with the Reflection Grating Spectrometer (RGS) and the
European Photon Imaging Camera (EPIC) onboard
the {\em XMM-Newton} satellite. TW\,Hya is the first classical T Tauri star for which
simultaneous X-ray data with both high spectral resolution and high sensitivity 
were obtained, thus allowing to probe the X-ray emission properties of stars in the early pre-main
sequence phase.  Despite  TW\,Hya's high X-ray luminosity in excess of $10^{30}$\,erg/s
its X-ray spectrum is dominated by emission lines from  rather cool plasma (T $\approx$ 3 MK), and only
little emission measure is present at high temperatures (T $\approx$ 10 MK).
We determine photon fluxes for the emission lines in the high resolution spectrum, 
confirming the earlier result from {\em Chandra}
that the predominant emission is from neon and oxygen, with comparatively weak iron lines.
Further, the line ratios of He-like triplets of nitrogen, oxygen and neon require  densities of
$n_{\rm e} \sim 10^{13}\,{\rm cm^{-3}}$, about two orders of magnitude higher than for any 
other star observed so far at high spectral resolution. Finally, we find that nearly all 
metals are underabundant with respect to solar abundances, while the
abundances of nitrogen and neon are enhanced.  The high plasma density, the 
(comparatively) low temperature, and peculiar chemical abundances in the 
X-ray emitting region on TW\,Hya are untypical for stellar coronae.  An alternative X-ray production 
mechanism is therefore called for and a natural explanation is an accretion column
depleted of grain forming elements. The metal depletion could be either due
to the original molecular cloud that formed TW\,Hya or due to a settling of dust in
the circumstellar disk of TW\,Hya.
}

\maketitle

\section{Introduction}\label{sect:intro}

T Tauri stars (TTS) are late-type pre-main sequence (PMS) stars,
usually found in or near star formation regions.
The two subclasses of TTS, termed `classical' and `weak-line', are
distinguished by the strength of their H$_\alpha$ emission.
This empirical distinction most likely also reflects different
evolutionary states: Most of the `classical' TTS (cTTS),
i.e., those with strong H$_\alpha$ emission, show signatures
of circumstellar matter
(e.g. IR and UV excesses) accreting onto the central star, while
in contrast, `weak-line' TTS (wTTS) show no signs for the presence of
disks which presumably have been lost.
A detailed description of early stellar evolution and the observable stellar
properties related with the different phases can
be found in \citey{Feigelson99.1}.

X-ray emission from TTS was first detected with the {\em Einstein}
satellite (e.g. \cite{Feigelson81.1}, \cite{Walter81.1}, \cite{Montmerle83.1}).
It was soon recognized that X-ray observations provide an efficient means of
discovering wTTS which can easily escape other identification methods because of their
lack of conspicuous features in the optical or near-infrared. Extensive X-ray studies
of a variety of star forming regions were undertaken with {\em ROSAT}. In particular
the spatial completeness of the {\em ROSAT} All-Sky Survey has resulted in a
large number of newly identified PMS stars
(e.g. \cite{Neuhaeuser95.1}, \cite{Alcala95.1}, \cite{Wichmann96.1}).

The origin of the X-ray emission from wTTS is commonly interpreted
in terms of scaled-up solar-type magnetic activity, but the situation is less clear
for their accreting counterparts, the cTTS.
While nothing should prevent the formation of magnetic fields and ensuing
coronae in cTTS, X-rays could alternatively also
be produced above the hot spot
where a magnetically funnelled accretion flow impacts on
the surface of the star (\cite{Lamzin99.1}).
A distinction between a corona or an accretion funnel as the site of
X-ray emission in cTTS requires a precise assessment
of the physical conditions in the X-ray emitting region.  With the lower
resolution X-ray data available from {\it Einstein-} and {\em ROSAT}-observations
of both cTTS and wTTS
it was impossible to adequately characterize the emission site,
although some differences between the two types of TTS were
indicated by statistical comparisons of their emission level and variability
(see e.g. \cite{Neuhaeuser95.1}, \cite{Stelzer00.1}, \cite{Stelzer01.1}).

With the advent of high-resolution X-ray spectroscopy with
the grating spectrometers onboard the {\em XMM-Newton} and
{\em Chandra} observatories this situation has fundamentally changed.
The reflection gratings onboard {\em XMM-Newton} and the transmission gratings
onboard {\em Chandra} have spectral resolutions sufficiently large to
resolve individual emission lines in the soft X-ray range. These emission
lines and in particular the analysis of
their relative strengths provide unprecedented tools for the plasma diagnostics
of the outer envelopes of stars and allow an assessment of the temperature structure,
density, and chemical abundances in the X-ray emitting region.

The high signal-to-noise ratio (SNR) required for a meaningful analysis
of high-resolution X-ray spectra clearly favors the study of the
nearest star forming regions.  The nearby TTS
TW\,Hya (located at a distance of only $\sim 57$\,pc; \cite{Wichmann98.1}) was
first thought to be an isolated cTTS (\cite{Rucinski83.1}), but later
five more stars were found in the same region and
shown to be members of a physical cluster of young stars dominated by TW Hya.
The whole group is now known as the TW~Hydra association (TWA; \cite{Kastner97.1}),
which is currently known to host about $20$ young stars
(youth being probed by Li absorption), most of which --
in contrast to TW\,Hya itself -- are wTTS.

The first -- and so far only -- example of a high-resolution X-ray spectrum
of an accreting PMS star was a {\em Chandra}/HETGS spectrum of
TW\,Hya presented by \citey{Kastner02.1}.  Interestingly, flux ratios of the OVII
triplett lines indicated densities much higher than found
in any stellar corona before, and consequently \citey{Kastner02.1}
called into question the solar-stellar analogy for TW\,Hya and
cTTS in general.  About a year after the {\em Chandra} observation
TW\,Hya was observed by {\em XMM-Newton}, resulting in a higher SNR
grating spectrum with the {\em XMM-Newton} Reflection Grating Spectrometer (RGS)
as well as a simultaneously measured
medium-resolution CCD spectrum with high
sensitivity over a broad energy range ($E = 0.3-10$\,keV) from the
European Photon Imaging Camera (EPIC).
The purpose of this paper
is to present the results of this 30\,ksec {\em XMM-Newton} observation
and properly discuss their implications in the context of X-ray emission
from TTS.

\section{Observations}\label{sect:observations}

TW\,Hya was observed with {\em XMM-Newton} on July 9, 2001 for a duration of
30\,ksec (Obs-ID 0112880201) with the RGS as prime instrument.
The observations
were performed in full-frame mode employing the medium filter for both the MOS and
pn cameras of EPIC. 
Because of the optical brightness of TW\,Hya ($V = 11$\,mag) the
Optical Monitor (OM) was in blocked position so that no concurrent optical
data is available. The observing log for all X-ray instruments onboard {\em XMM-Newton}
is given in Table~\ref{tab:obslog}.
%
%
\begin{table}
\begin{center}
\caption{Observing log for the {\em XMM-Newton} observation of TW\,Hya on July 9, 2001. "Start" and "Stop" refer to start and end of exposures.}
\label{tab:obslog}
\begin{tabular}{lrrrrr}\hline
Instr. & \multicolumn{2}{c}{UT [hh:mm:ss]} & \multicolumn{2}{c}{JD - 2452099} & Expo \\
      & \multicolumn{1}{c}{Start} & \multicolumn{1}{c}{Stop} & \multicolumn{1}{c}{Start} & \multicolumn{1}{c}{Stop} & \multicolumn{1}{c}{[ksec]} \\ \hline
pn    & 06:35:38 & 13:57:31 & 0.774306 & 1.081250 & 26.52 \\
MOS   & 05:58:02 & 14:01:48 & 0.748611 & 1.084028 & 28.98 \\
RGS   & 05:51:39 & 14:05:15 & 0.743750 & 1.086806 & 29.64 \\ \hline
\end{tabular}
\end{center}
\end{table}

TW\,Hya is by far the brightest X-ray source
and the only known member of the TW\,Hya association in this EPIC field.
By inspecting the time-resolved background emission we
verified that the background was low and showed only little time
variability without any significant solar-activity related contamination, so that
the full observing time provided data useful for analysis.
The pn data shows slight evidence for pile-up. Therefore, we base our analysis
on the MOS detectors, which furthermore provide a better energy resolution.

\begin{figure}
\begin{center}
\resizebox{9cm}{!}{\includegraphics{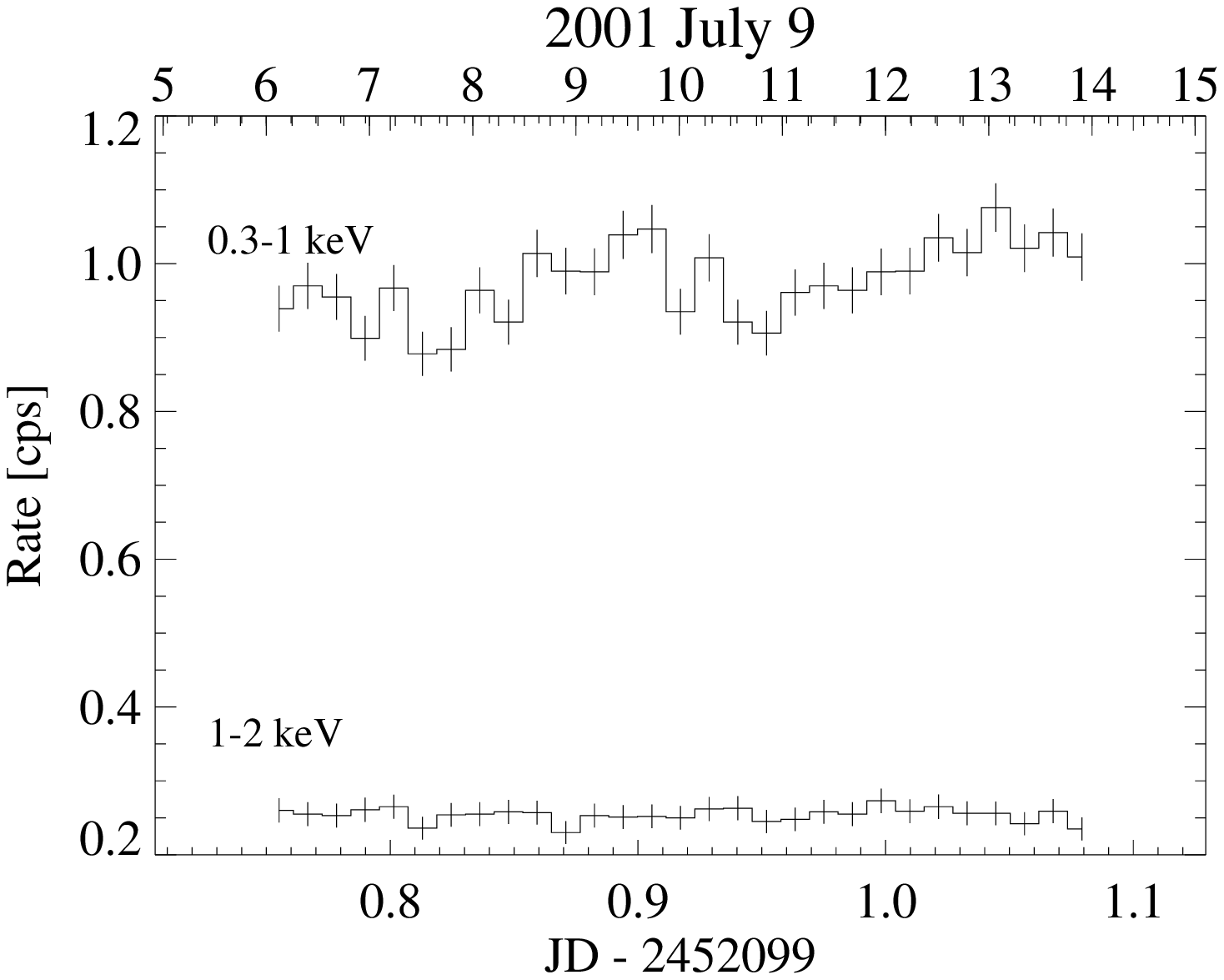}}
\resizebox{9cm}{!}{\includegraphics{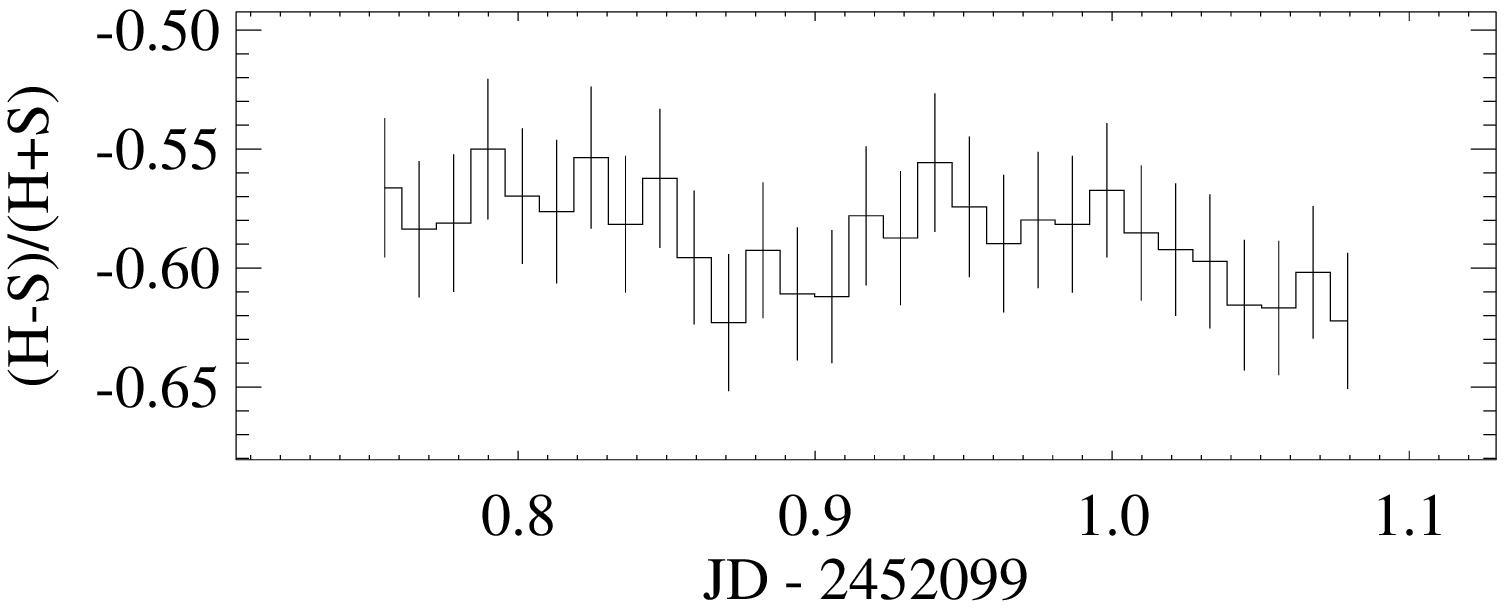}}
\caption{Background subtracted MOS lightcurves of TW\,Hya in the energy range $0.3-1.0$\,keV and $1.0-2.0$\,keV. The panel on the bottom shows the hardness ratio computed from these bands; binsize is 1\,ksec.}
\label{fig:epic_lc}
\end{center}
\end{figure}

We constructed an X-ray light curve of TW\,Hya by binning all photons from a
circular area around the target position.
We used an extraction radius of $40^{\prime\prime}$ including about 90\,\% of the
source photons. For the background
subtraction we used a source-free circle of the same area located near TW\,Hya.
Background subtracted MOS lightcurves of TW\,Hya in the energy range
$0.3-2$\,keV are shown in Fig.~\ref{fig:epic_lc}. The counts collected by MOS\,1
and MOS\,2 have been added to increase the SNR.

To investigate spectral changes during the observation
we defined two energy bands, $S$ for `soft' and $H$ for `hard', and
computed a hardness ratio $HR = (H-S)/(H+S)$. 
The overall X-ray spectrum of TW\,Hya is very soft, with
only a few photons recorded above $\sim 2$\,keV (see discussion
below). To ensure sufficient signal
in both bands we chose $S = 0.3-1.0$\,keV and $H=1.0-2$\,keV.
The time evolution of $HR$ is shown in the lower panel of
Fig.~\ref{fig:epic_lc}. 
Neither the lightcurve nor the evolution of the $HR$ show conspicuous
signs for flaring. 
%
%

\section{High-resolution Spectrum: RGS}\label{sect:rgs}

\begin{figure*}
\begin{center}
\hspace*{-1.cm}\resizebox{20cm}{!}{\includegraphics{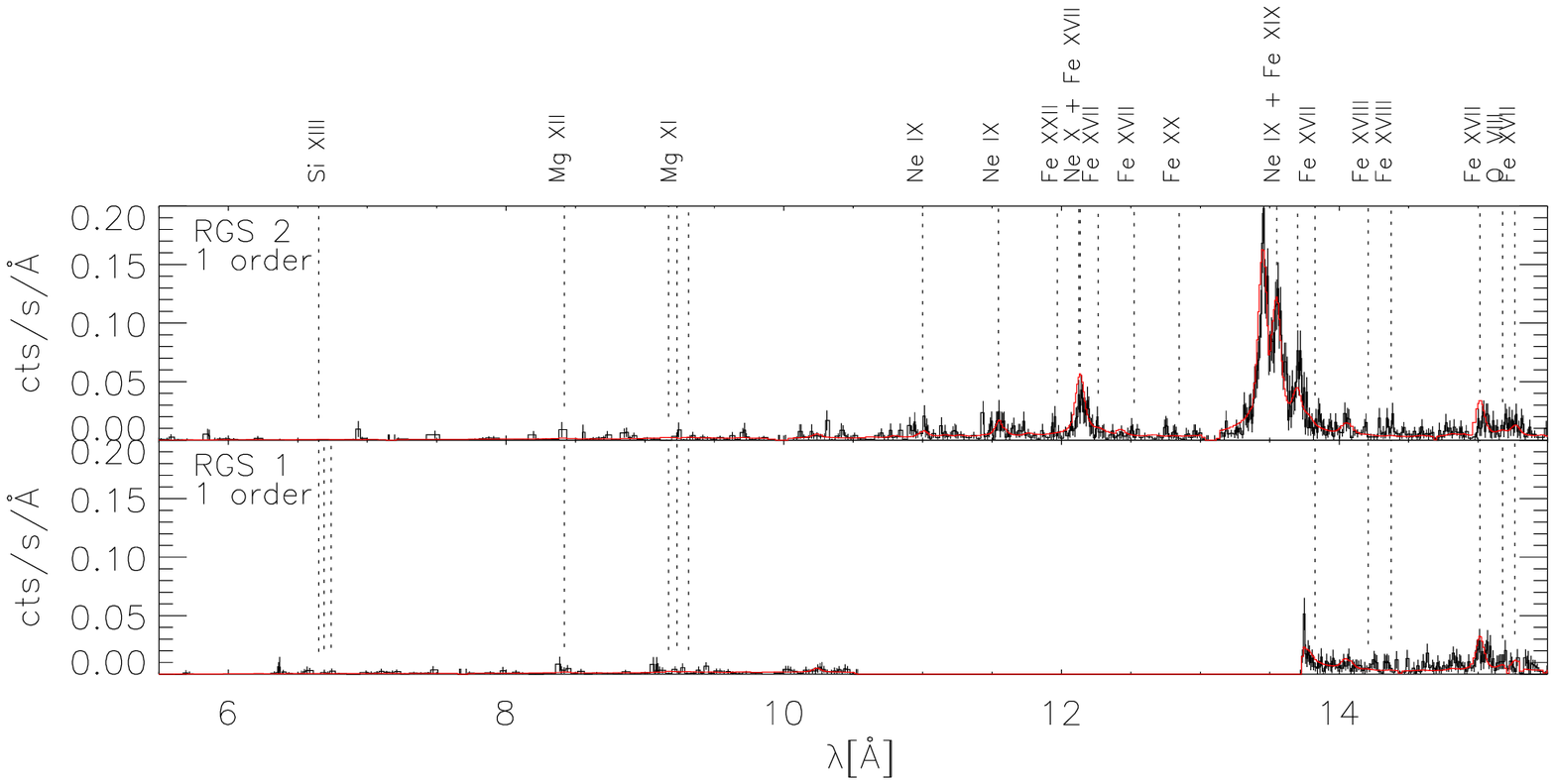}}
\hspace*{-1.cm}\resizebox{20cm}{!}{\includegraphics{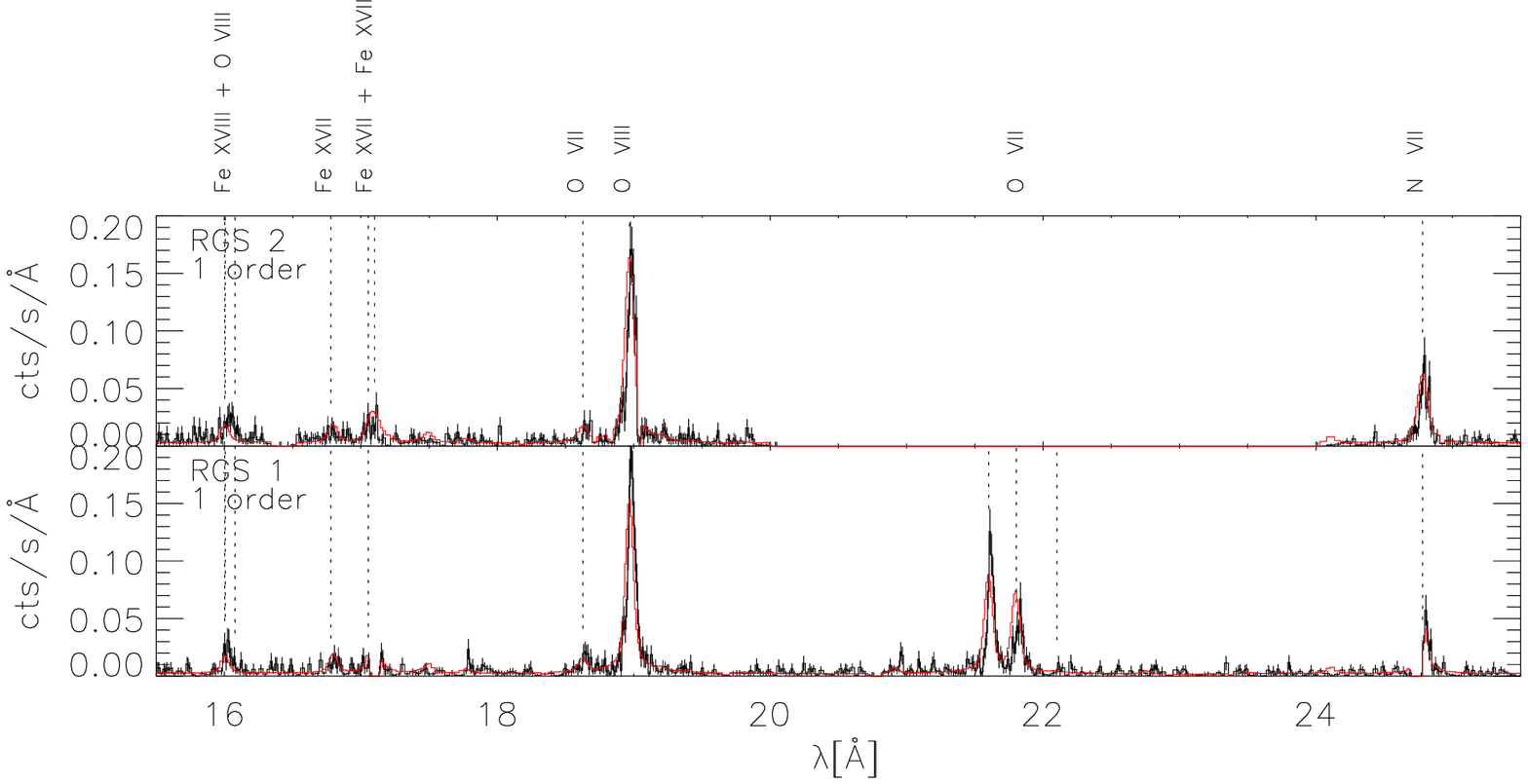}}
\hspace*{-1.cm}\resizebox{20cm}{!}{\includegraphics{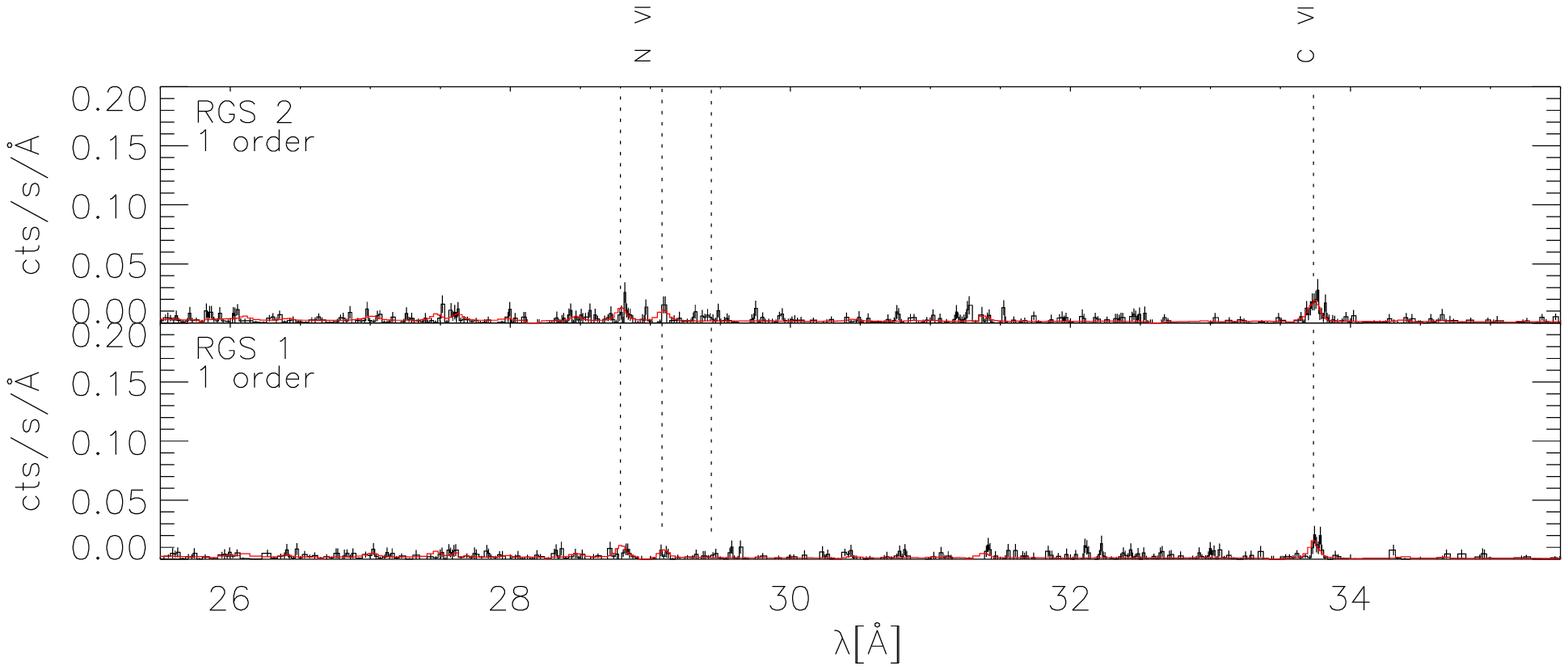}}
\caption{First order background subtracted {\em XMM-Newton} RGS count rate spectrum of TW\,Hya.
Overplotted is a 3-T VMEKAL model whose parameters were derived from a combination of emission line
analysis and global fitting of the medium-resolution MOS spectrum. The MOS spectrum is shown together with the same model in Fig.~\ref{fig:mos}, and the best fit parameters are summarized in Table~\ref{tab:spec_params}. Exposure time is 29\,ksec for each RGS. Straight horizontal lines represent gaps due to CCD chain failure or individual chip separation. Emission lines typical for stellar coronae are indicated by labels and dashed lines.}
\label{fig:rgs}
\end{center}
\end{figure*}

The RGS data of TW\,Hya were analysed using the standard pipeline {\it rgsproc}
of the {\em XMM-Newton} Science Analysis System (SAS), version 5.4.1.
We extracted the total (source plus background) and a background spectrum
for TW\,Hya. In this paper only the first dispersion order is discussed.
The full first order RGS spectrum of TW\,Hya is displayed
in Fig.~\ref{fig:rgs}, except for
the small range between $35-37$\,\AA~ where no spectral lines were found.
In Fig.~\ref{fig:rgs} identifications of the stronger emission lines
are also provided from comparisons with line lists of the MEKAL
code (\cite{Mewe85.1}, \cite{Mewe95.1}), the National Institute of Standards
(NIST) data base, and the results of solar flare observations
(\cite{Phillips99.1}).  In the RGS spectrum one recognizes the lines of H-like
and He-like carbon, nitrogen, oxygen and neon, while iron lines are weak
or absent (especially those near 15 and 17 \AA ). The typical
silicon and magnesium lines below 10\,\AA~ are also absent in the RGS spectrum.
Note, however, that at those wavelengths the RGS effective area decreases rapidly.

\subsection{Identification and Analysis of Emission Lines}\label{subsect:rgs_lines}

To determine fluxes for individual emission lines we used the
CORA\footnote{CORA can be downloaded from
http://www.hs.uni-hamburg.de/DE/Ins/Per/Ness/Cora} line fitting tool.
CORA is based on a maximum likelihood technique (see \cite{Ness02.2}) and
fits a specified number of emission lines combined with a vertical offset
to account for a constant background (both instrumental and due to
continuum emission from the source) assuming some specified
emission line profile function.
Earlier studies have shown that a Lorentzian profile provides
a reasonable description of the shape of RGS line profiles
(e.g. \cite{Ness02.1}, \cite{Stelzer02.1}).
Before feeding the spectra into CORA we corrected for the different extraction
area of total and background spectrum.
To take into account possible
small background variations we selected successive wavelength intervals
in which the background can be approximated by a constant and performed the analysis
separately for each of these intervals.
Some ambiguity does remain in the determination of the local background
level. To account for uncertainties associated with the background level
we add a systematic error to the statistical error of the line counts.
This systematic error was derived by using the highest and lowest background
levels that would provide an acceptable line fit after visual inspection of the
solution. Finally, the line counts and their errors (both statistical and systematic)
were transformed into photon fluxes with the help of the effective areas of the RGS.

A list of the brightest lines in the spectral range of the RGS
($6-37$\,\AA) is provided in Table~\ref{tab:rgs_lines}, where we list the
lines' central wavelengths and their errors, the fitted number of counts in the
line including errors and
the derived photon fluxes. A suggested identification is also given.
A substantial part of the total wavelength
range is covered only by one instrument due to failure of one CCD on each RGS, and
because of the gaps between adjacent CCDs.
For lines in these regions we give the result of the active RGS, for the lines covered
by both RGS we summed up the two spectra.
For diagnostic purposes the most important lines are
the $Ly_\alpha$ and triplet lines of hydrogen-like and helium-like ions.  The
measurements of these lines for all species detected in the RGS spectrum and in addition
the most prominent iron lines are provided in Table~\ref{tab:rgs_lines}.
We emphasize again the unusual weakness of iron lines
as one of the most remarkable features of the X-ray spectrum of TW\,Hya.

%
%
%

%
%
\begin{table*}
\begin{center}
\caption{Spectral lines identified in the first order {\em XMM-Newton} RGS spectrum of TW\,Hya. {\em left} - Observed line center, width, number of counts, and photon flux. Uncertainties include statistical 1$\sigma$ error plus a systematic error stemming from the uncertainty of the local background level (see text for details). {\em right} -  Line identifications from \protect\citey{Mewe85.1}, \protect\citey{Mewe95.1}, and \protect\citey{Phillips99.1}. For line blends we list the most likely contributors according to X-ray observations of other stars.
}
\label{tab:rgs_lines}
\begin{tabular}{rrrr|llr} \hline
\multicolumn{1}{c}{$\lambda$} & \multicolumn{1}{c}{$\sigma$} & \multicolumn{1}{c}{$I$} & \multicolumn{1}{c|}{Photon Flux} & \multicolumn{3}{c}{Identification} \\
\multicolumn{1}{c}{[\AA]} & \multicolumn{1}{c}{[\AA]} & \multicolumn{1}{c}{[cts]} & \multicolumn{1}{c|}{[$10^{-5}\,\frac{\rm ph}{\rm s\,cm^2}$]} & Ion & Trans. & $\lambda$ \\
\hline
\multicolumn{7}{c}{\bf RGS\,1} \\
\hline
$18.983$ & $0.055$  & $465.7 \pm 43.7$ & $30.4 \pm 2.8$ & O\,VIII & $Ly_\alpha$ & 18.97 \\
$21.616$ & $0.047$  & $241.6 \pm 28.4$ & $17.4 \pm 2.1$ & O\,VII & $r$ & 21.61 \\
$21.836$ & $0.053$  & $128.2 \pm 20.1$ & $ 9.5 \pm 1.5$ & O\,VII & $i$ & 21.80 \\
$22.100$ & $0.037$  & $  5.3 \pm  7.1$ & $ 0.4 \pm 0.6$ & O\,VII & $f$ & 22.10 \\
\hline
\multicolumn{7}{c}{\bf RGS\,2} \\
\hline
$12.152$ & $0.060$ & $111.4 \pm 26.1$ & $ 6.3 \pm 1.5$ & Ne\,X    & $Ly_\alpha$ & 12.13 \\
$13.460$ & $0.050$ & $402.0 \pm 38.1$ & $23.1 \pm 2.2$ & Ne\,IX & $r$ & 13.45 \\
$13.560$ & $0.057$ & $308.9 \pm 19.9$ & $17.4 \pm 1.1$ & Ne\,IX & $i$ & 13.55 \\
$13.710$ & $0.054$ & $149.7 \pm 27.3$ & $ 8.2 \pm 1.5$ & Ne\,IX & $f$ & 13.70 \\
$17.065$ & $0.075$ & $ 46.8 \pm 15.9$ & $ 2.7 \pm 0.9$ & Fe\,XVII & $Ne3G$ & 17.06 \\
$24.793$ & $0.060$ & $171.5 \pm 31.5$ & $11.8 \pm 2.2$ & N\,VII  & $Ly_\alpha$ & 24.78 \\
$28.820$ & $0.059$ & $ 32.1 \pm 14.2$ & $ 2.9 \pm 1.3$ & N\,VI   & $r$ & 28.79 \\
$29.101$ & $0.030$ & $ 14.2 \pm  7.2$ & $ 1.3 \pm 0.6$ & N\,VI   & $i$ & 29.10 \\
$29.540$ & $0.030$ & $  4.3 \pm  5.8$ & $ 0.4 \pm 0.5$ & N\,VI   & $f$ & 29.54 \\
\hline
\multicolumn{7}{c}{\bf RGS\,1 + RGS\,2} \\
\hline
$16.030$ &  $0.056$ & $118.7 \pm 37.2$ & $ 3.5 \pm 1.1$ &  Fe\,XVIII & & 16.00 \\
         &         &                  &                 &  O\,VIII & $Ly_\beta$ &$16.01$ \\
$18.649$ &  $0.037$ & $ 53.4 \pm 12.1$ & $ 1.7 \pm 0.4$ &  O\,VII  & $He3A$ & 18.63 \\
$33.757$ &  $0.065$ & $113.9 \pm 20.7$ & $ 7.8 \pm 1.4$ &  C\,VI   & $Ly_\alpha$ & 33.70 \\
\hline
\end{tabular}
\end{center}
\end{table*}

\subsection{Line Temperatures}\label{subsect:line_temp}

\begin{figure}
\begin{center}
\parbox{9cm}{\resizebox{9cm}{!}{\includegraphics{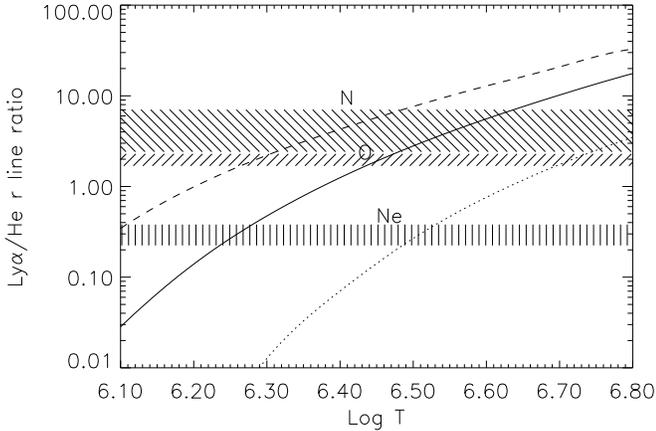}}}
\caption{Temperature dependence of the $Ly_\alpha/r$ line ratio for neon (dotted line),
oxygen (solid line), and nitrogen (dashed line) according to the CHIANTI code 
(\protect\cite{Dere97.1}). The ratios observed in the RGS spectrum of TW\,Hya are
indicated by the shaded patterns. All observed values are compatible with a $2-3$\,MK
hot isothermal plasma.}
\label{fig:temp_ratios}
\end{center}
\end{figure}

For the elements neon, oxygen and nitrogen both the $Ly_\alpha$ emission lines at
12.1\,\AA, 19.0\,\AA, and 24.8\,\AA \ respectively, and the He-like resonance lines at
13.5\,\AA, 21.6\,\AA, and 28.8\,\AA \ respectively
are detected in our RGS spectra.
In Fig.~\ref{fig:temp_ratios} we plot the
expected energy flux ratio from the H-like $Ly_\alpha$ and He-like resonance
lines for neon (dotted line), oxygen (solid line) and nitrogen (dashed line) versus
logarithmic temperature using the CHIANTI calculations (\cite{Dere97.1}). 
We also indicate the measured line ratio range for
neon (upright shading), oxygen (right tilted shading) and nitrogen (left tilted shading).
These line ratios clearly correspond to effective ``isothermal'' temperatures, which, however,
are independent of abundance. For
oxygen and nitrogen these temperatures agree at $\lg{T}\,[K] \approx 6.45$, 
for neon a slightly larger effective temperature is found.
The X-ray spectrum of TW\,Hya does not contain any strong high temperature lines
despite its large overall X-ray luminosity. Thus we conclude that most of the observed line
emission seems to be produced by material located at pretty much the same temperature.
In fact, we verified that a 1-T thermal model (VMEKAL) with free abundances can provide
an acceptable description of the RGS spectrum ($\chi^2_{\rm red} = 1.36$ for $2333$ d.o.f.).

However, one needs to keep in mind that the RGS effective area decreases rapidly at
wavelengths below $\sim 10$\,\AA, i.e. energies above 1.2\,keV.
Elements such as Mg and Si have their characteristic He-like and
H-like line transitions in this wavelength region. In the RGS spectrum of TW\,Hya these
lines remain undetected because of lacking SNR, but this
critical spectral region can be studied with EPIC at medium resolution
(see Sect.~\ref{sect:epic}). The EPIC spectrum does also show a high-energy tail that
can not be explained by the rather low plasma temperatures dominating the RGS line
emission (cf., Sect.~\ref{sect:epic} and Fig.~\ref{fig:mos}). 

\subsection{Differential Emission Measure}\label{subsect:dem}

In order to quantify a differential emission measure (DEM) distribution we adopted the
following approach; see also \cite{Schmitt04.1} for more details on this DEM reconstruction 
method. 
We first introduced a fixed temperature grid located at grid points
log T = [6.3,6.5,7.3]. The two lower temperatures are based on the result from the above
described line analysis, while the highest temperature is required for a match to the
medium-resolution EPIC spectrum whose high-energy tail cannot possibly be
described with a 3\,MK plasma alone. It turns out, however, that the contribution
of the hot component to the overall RGS count rate is very small.

At each temperature grid point $T_i$ (i=1,3) we assume the presence of some emission measure
$EM_i$.  Denoting the elements carbon (j=1), nitrogen (j=2), oxygen (j=3) and
neon (j=4) with the index j, we can then compute the line emissivities $P_{\rm H,i,j}$ and
$P_{\rm He,i,j}$ for the hydrogen- and helium-like lines of the elements j for the temperature
components i.
The observed line ratios $\rho_{\rm obs,j}$ (j=2,4) can then be modelled as
\begin{equation}
\rho_{\rm mod,j} = \frac {\sum_{\rm i=1}^{3} EM_i \cdot P_{\rm H,i,j}} {\sum_{\rm i=1}^{3} EM_i \cdot P_{\rm He,i,j}}
\end{equation}
and we seek a best fit set of the emission measures $EM_i$ through minimizing
\begin{equation}
\chi^2 = \sum_{\rm j=2}^{4} \frac {(\rho_{\rm obs,j} - \rho_{\rm mod,j})^2} {(\sigma_{\rm j})^2},
\end{equation}
where $\sigma_{\rm j}$ denotes the error of the line ratio j. Clearly, since we are dealing with line
ratios of the same elements, the normalization can be set arbitrarily (we chose $EM_1$ = 1).
The thus derived EM-distribution is abundance independent. The best fit values
for the (relative) values of $EM_{\rm i}$ are found to be [1.0,2.35,0.59]. This choice of emission measure
reproduces the observed line ratios extremely well ($\chi^2_{\rm red} \approx 0.5$). However, given
the ill-conditioned nature of the reconstruction problem, other choices of the EM-distribution
also provide statistically acceptable fits. With the thus derived set of best fit
$EM_i$ values we can choose the normalization to reproduce, say, the oxygen $Ly_{\alpha}$ line
fluxes. Then we use this normalization to determine abundances of carbon, nitrogen, neon and iron
relative to oxygen by forcing agreement between predicted and measured fluxes in the observed lines.
Note that for carbon and iron we can only use the $Ly_{\alpha}$ line (for carbon) and the
Fe XVII 17.07~\AA~ line (for iron). Using an interstellar hydrogen column density of
$N_H = 2 \times 10^{20}$ cm$^{-2}$ -- which is required for a good fit of the EPIC data and is in 
agreement with the expected interstellar absorption towards TW\,Hya -- 
we find with this procedure the values  
$C/O=1.07$, $N/O = 3.15$, $Ne/O = 9.77$ and $Fe/O = 0.91$ with respect to cosmic abundances.
Realistic abundance errors are difficult to determine. We estimate abundance errors 
of about 20\% for neon, and about 30 \% for carbon, nitrogen and iron.  

\begin{table} 
\begin{center}
\caption{Best fit parameters for the X-ray spectrum of TW\,Hya described by a multi-temperature thermal model (VMEKAL) including photo-absorption ($N_{\rm H} = 2 \times 10^{20}\,{\rm cm^{-2}}$). The electron density was fixed to $5 \times 10^{12}\,{\rm cm^{-3}}$. Elements not listed below are irrelevant in this spectrum, and their individual abundances were held fixed at the solar value. The temperatures and the coupling between some parameters were pre-determined by the analysis of the emission lines seen in the RGS spectrum. Values for the abundances are given with respect to the solar photosphere, $\log{T}$ is in K, and $\log{EM}$ in ${\rm cm^{-3}}$. (A) - Best fit 3-T model to the EPIC MOS spectrum in the range $\lambda = 5 - 40$\,\AA. (B) - Same as (A) with free Mg and Si abundances.}  
\label{tab:spec_params}
\begin{tabular}{lrrrr} \hline
          & \multicolumn{2}{c}{---------Model~A---------} & \multicolumn{2}{c}{---------Model~B---------} \\ \hline
          & param.   & param. & param.  & param.\\
Parameter & coupling & value & coupling & value \\
\hline
$C$       &  $=1.07 \times O$ & $=0.28$ & $=1.07 \times O$ & $0.40$ \\
$N$       &  $=3.14 \times O$ & $=0.82$ & $=3.14 \times O$ & $1.18$ \\
$O$       &                   & $0.26$  &                  & $0.37$ \\
$Ne$      &                   & $3.15$  &                  & $3.97$ \\
$S$       &                   & $=0.0$  &                  & $=0.0$ \\
$Mg$      &                   & $=1.0$  &                  & $0.65$ \\
$Si$      &                   & $=1.0$  &                  & $0.13$ \\
$Fe$      &                   & $0.26$  &                  & $0.24$ \\
\hline
$\log{T_1}$    & & $=6.3$ & & $=6.3$ \\
$\log{T_2}$    & & $=6.5$ & & $=6.5$ \\
$\log{T_3}$    & & $=7.3$ & & $=7.3$ \\
\hline
$\log{EM_1}$    &                     & $52.64$ & & $52.53$ \\
$\log{EM_2}$    &  $2.35 \times EM_1$ & $=53.01$ & $2.35 \times EM_1$ & $=52.90$ \\
$\log{EM_3}$    &  $0.59 \times EM_1$ & $=52.41$ & & $52.53$ \\
\hline
$\chi^2_{\rm red}$ (d.o.f.) & \multicolumn{2}{c}{$2.79$ ($233$)} & \multicolumn{2}{c}{$2.61$ ($232$)} \\
\hline
\end{tabular}
\end{center}
\end{table}

\section{EPIC medium resolution spectrum: EPIC}\label{sect:epic}

The count rate spectrum of each EPIC MOS detector was extracted from the area
described in Sect.~\ref{sect:observations}.
We extracted and analysed the EPIC spectra with SAS version 5.4.1.
Individual redistribution matrices 
and an ancilliary response file were generated in the course of our data
reduction process. The combination of these two files fully models the 
instrumental response. 

The EPIC spectrum has a pronounced peak at $E \sim 0.91$\,keV or $13.6$\,\AA.
Since we know from the {\em XMM-Newton} RGS spectrum that both the continuum and
iron line emissions are weak,  we conclude that this spectral bump must
represent emission from H- and He-like neon.
This remarkable feature is not observed in  EPIC data of other young stars
(see e.g. \cite{Guedel01.1}, \cite{Stelzer02.1}, \cite{Stelzer01.2}),
emphasizing the peculiar nature of the X-ray spectrum of TW\,Hya.

Approximating the MOS spectra by a sum of three thermal models with
fixed values of temperature and
based on the parameters derived from the line analysis, we
verify that the temperature and abundance structure derived
from the analysis of individual emission lines also provides a consistent
description of the medium-resolution data. 
In this global fit to the MOS spectrum
an absorption term with atomic cross-sections of \citey{Morrison83.1}
and $N_{\rm H} = 2 \times 10^{20}\,{\rm cm^{-2}}$ is applied to describe the
effects of inter- and/or circumstellar cold gas.
The electron density was set to $5 \times 10^{12}\,{\rm cm^{-3}}$ 
following the result of our line-based density analysis  
(see Sect.~\ref{subsect:triplets}) and
kept fixed. We restricted our modeling to the $5.5 - 40$\,\AA~ range,
and used the VMEKAL code as implemented in XSPEC, version 11.2.0.

%
%
\begin{figure}
\begin{center}
\parbox{9cm}{\resizebox{9cm}{!}{\includegraphics{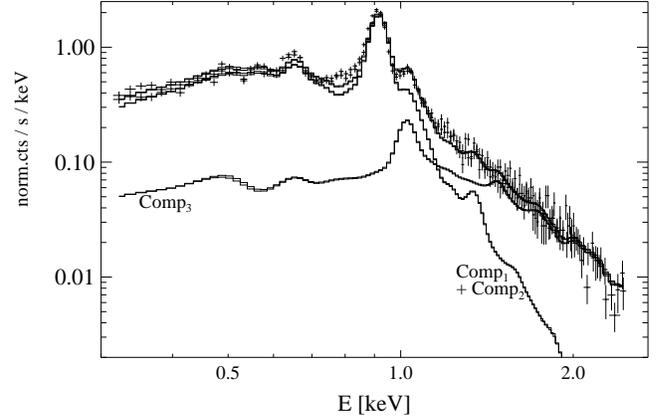}}}
\caption{EPIC MOS\,1 and MOS\,2 spectrum of TW\,Hya. The data is overlaid by the best fit 3-T VMEKAL model~B from Table~\ref{tab:spec_params}; see also Fig.~\ref{fig:rgs} where the same model is folded with the RGS response. The individual contributions of the two lower-temperature components and of the high-temperature component are also shown.} 
\label{fig:mos}
\end{center}
\end{figure}

We first allowed only the oxygen abundance and the absolute value of the emission measure
to vary, i.e., the parameters not constrained by our line modeling.
The best fit determined in this fashion results in $\chi^2_{\rm  red} = 3.2$ (235 d.o.f.),
and the residuals suggest that a slight adjustment of the abundances may be needed. 
If we allow the neon and
iron abundances to de-couple from the oxygen abundance, the fit improves to
$\chi^2_{\rm red} = 2.8$ (233 d.o.f.). Note that these ``new'' abundances are still within the
errors of the abundances derived from our line-based RGS analysis.
The best fit parameters of this model are summarized in Table~\ref{tab:spec_params}\,(Model\,A).
We point out that we do not derive nitrogen  and carbon abundances from the EPIC spectra,
because the instrument has little sensitivity in the relevant spectral range. These abundances
are solely derived from the RGS data and -- for carbon -- are based only on one single line.

Up to this point our fit results are in good
agreement with the findings presented by \citey{Kastner02.1} from their
analysis of the {\em Chandra} HETG spectrum of TW\,Hya, i.e., a strong over-abundance of Ne
and emission dominated by low-temperature plasma.
Inspection of the EPIC fit residuals does however show clear discrepancies
between the EPIC data and model at $\lambda < 14$\,\AA,
where the H-like $Ly_\alpha$ and He-like triplet
transitions of magnesium and silicon are located.
In particular, two bumps in the model come from emission of Si\,XIV and Mg\,XII without
any counterpart in the data.  Therefore we allowed for adjustment of the Mg and Si abundances
in the spectral fitting process. Since any changes in the Mg and Si
abundances will be compensated for by changes in $EM_3$, we must also let vary $EM_3$. The
resulting best fit models have magnesium and silicon abundances well
below solar values with a best-fit $\chi ^2_{\rm red}$ of 2.61. The 90\,\% error contour of the
magnesium is at 0.94 (w.r.t. solar), and the silicon abundance is definitely subsolar.
A summary of all parameters of this model is given in Table~\ref{tab:spec_params} (Model\,B).
In Fig.~\ref{fig:mos} we show the best fit 3-T model together with the MOS data.
For illustration the sum of the two low-energy components and the high-energy component
are also plotted separately.

\begin{figure}
\begin{center}
\parbox{9cm}{\resizebox{9cm}{!}{\includegraphics{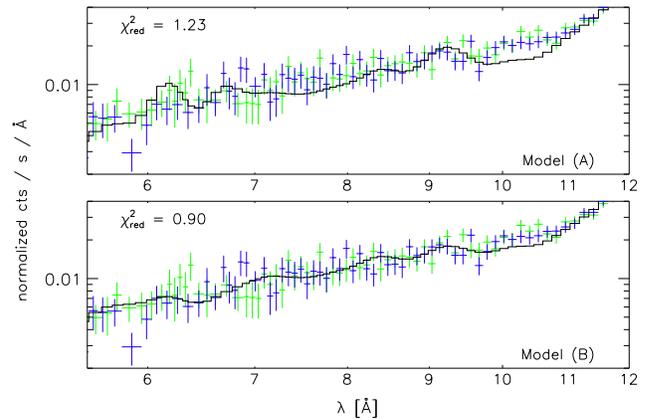}}}
\caption{Close-up of the EPIC MOS\,1 and MOS\,2 spectrum of TW\,Hya in the range $5.5 - 12$\,\AA: {\em top} -  Model (A) from Table~\ref{tab:spec_params}: Mg and Si abundance fixed to solar; {\em bottom} - Model (B) from Table~\ref{tab:spec_params}: Mg and Si abundance free parameters resulting in subsolar abundances for both elements.}
\label{fig:mos_2models}
\end{center}
\end{figure}

In order to point out the effects of changing Si and Mg abundances we consider
only the range between $5.5 - 12$\,\AA.
In Fig.~\ref{fig:mos_2models} we confront the two models from Table~\ref{tab:spec_params}
in this wavelength interval with each other.
Clearly the fit with free abundances describes the EPIC data far better
than assuming solar abundances for silicon and magnesium.
In the $5.5 - 12$\,\AA~ interval the
$\chi^2_{\rm red}$ decreases from $1.23$ ($102$ d.o.f.) for the model with solar
Mg and Si to $0.90$ ($101$ d.o.f.) for the model where Mg and Si are free parameters,
leading to subsolar abundances for both elements.

As a final consistency check of our procedure we now folded back the best fit
model~B with the response of the RGS. The result is overplotted on the RGS data in
Fig.~\ref{fig:rgs}. As expected, we find good agreement between model and data.
The emission measure of the high-T component is about a factor of
three lower than the sum of the two low-T components. This together with its high
temperature leads to a very low contribution to the RGS count rate of 0.015\,cps (RGS\,1) and
0.021\,cps (RGS\,2), implying that the high-T component is essentially undetectable in the
RGS. (The lower count rate in RGS\,1 is caused by the missing CCD in the $10-14$\,\AA~ interval.)
We thus conclude that the model~B described in Table~\ref{tab:spec_params} provides an acceptable
description of both the {\em XMM-Newton} EPIC and RGS spectra.

%

\section{Discussion}\label{sect:discussion}

So far TW\,Hya is the only cTTS, i.e. a PMS star with
strong H$_\alpha$ emission,
for which a high-resolution {\em XMM-Newton} X-ray spectrum is
available. Since in young stars H$_\alpha$ emission is related to accretion
TW\,Hya presents a unique test case for X-ray emission in this early
phase of stellar evolution.
Furthermore, it is one of the few
stars observed with both {\em XMM-Newton} and {\em Chandra} at high spectral
resolution.
In the following we discuss the properties of the X-ray emission of  TW\,Hya, and
we present a physical scenario explaining it.

\subsection{Variability}\label{subsect:variability}

During the {\em Chandra} observation of TW\,Hya discussed by
\citey{Kastner02.1} a sharp rise in intensity followed by a nearly
linear decay occurred, that the authors interpreted as a flare.
We find little evidence for variability 
in the {\em XMM-Newton} observation presented here 
(acquired $\sim$\,1\,yr after the {\em Chandra} data). 
In any case the lightcurve 
does not carry the clear signatures of a flare, 
and no change in spectral hardness is observed. 
Similarly \citey{Kastner02.1} found no changes in the
spectral parameters during the variable phase of the {\em Chandra}
observation, very untypical for stellar (flare) activity.
The flare-like event in the
{\em Chandra} observation has a decay time of $\sim 10$\,ksec. Assuming
a flare loop at $T \sim 3$\,MK (the dominating temperature derived from
the X-ray spectrum) that decays through radiative cooling the density
should be $\sim 10^{10}\,{\rm cm^{-3}}$. This is in sharp contrast to the
result from the analysis of line ratios (see Sect.~\ref{subsect:triplets}),
according to which the densities are higher by two orders of magnitude.

We directly compare the emission levels by computing X-ray fluxes for the
time-averaged spectrum. From the {\em Chandra} HETGS spectrum
\citey{Kastner02.1} derived an X-ray flux of
$2.5\times10^{-3}\,{\rm photons\,cm^{-2}\,s^{-1}}$
for the $0.45-6.0$\,keV band.
Integrating the RGS spectrum in the energy band which comes nearest
to the spectral range of the HETGS,
i.e. $5.5 - 27.5$\,\AA\ or $0.45-2.25$\,keV,
we arrive at a similar value of
%
%
$f_{\rm x,RGS\,1} \approx f_{\rm x,RGS\,2} = 2.3\times10^{-3}\,{\rm photons\,cm^{-2}\,s^{-1}}$
corresponding to $L_{\rm x,RGS} = 1.4\times10^{30}\,{\rm erg/s}$
at a distance of $57$\,pc.
This suggests that TW\,Hya exhibits only moderate long-term variability.

\begin{figure}
\begin{center}
\resizebox{9cm}{!}{\includegraphics{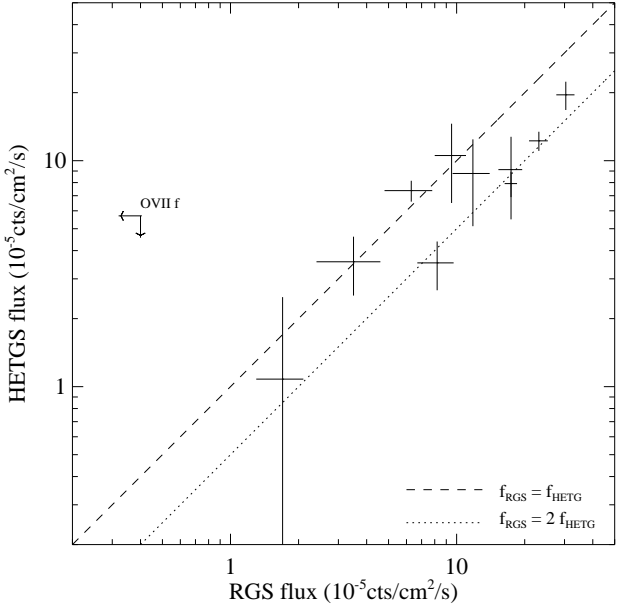}}
\caption{Comparison of photon fluxes for individual emission lines in the X-ray spectrum of TW\,Hya measured with the {\em XMM-Newton} RGS (see Table~\ref{tab:rgs_lines})
and the {\em Chandra} HETGS-MEG (see \protect\cite{Kastner02.1}). The {\em XMM-Newton} and {\em Chandra} observations are separated by about one year. All lines are constant within about a factor of two.}
\label{fig:fluxcomp}
\end{center}
\end{figure}

Based on the small amplitude of long-term variations discussed above,
comparing fluxes measured for individual emission lines allows to check
the calibration of the instruments on both satellites. We observe a trend
towards slightly higher fluxes in the RGS as compared to the
HETGS (Fig.~\ref{fig:fluxcomp}).
However, the difference is a factor two at most.
%
%

\subsection{Density in the X-ray Emitting Region}\label{subsect:triplets}

%
%
\begin{figure*}
\begin{center}
\parbox{18cm}{
\parbox{9cm}{\resizebox{9cm}{!}{\includegraphics{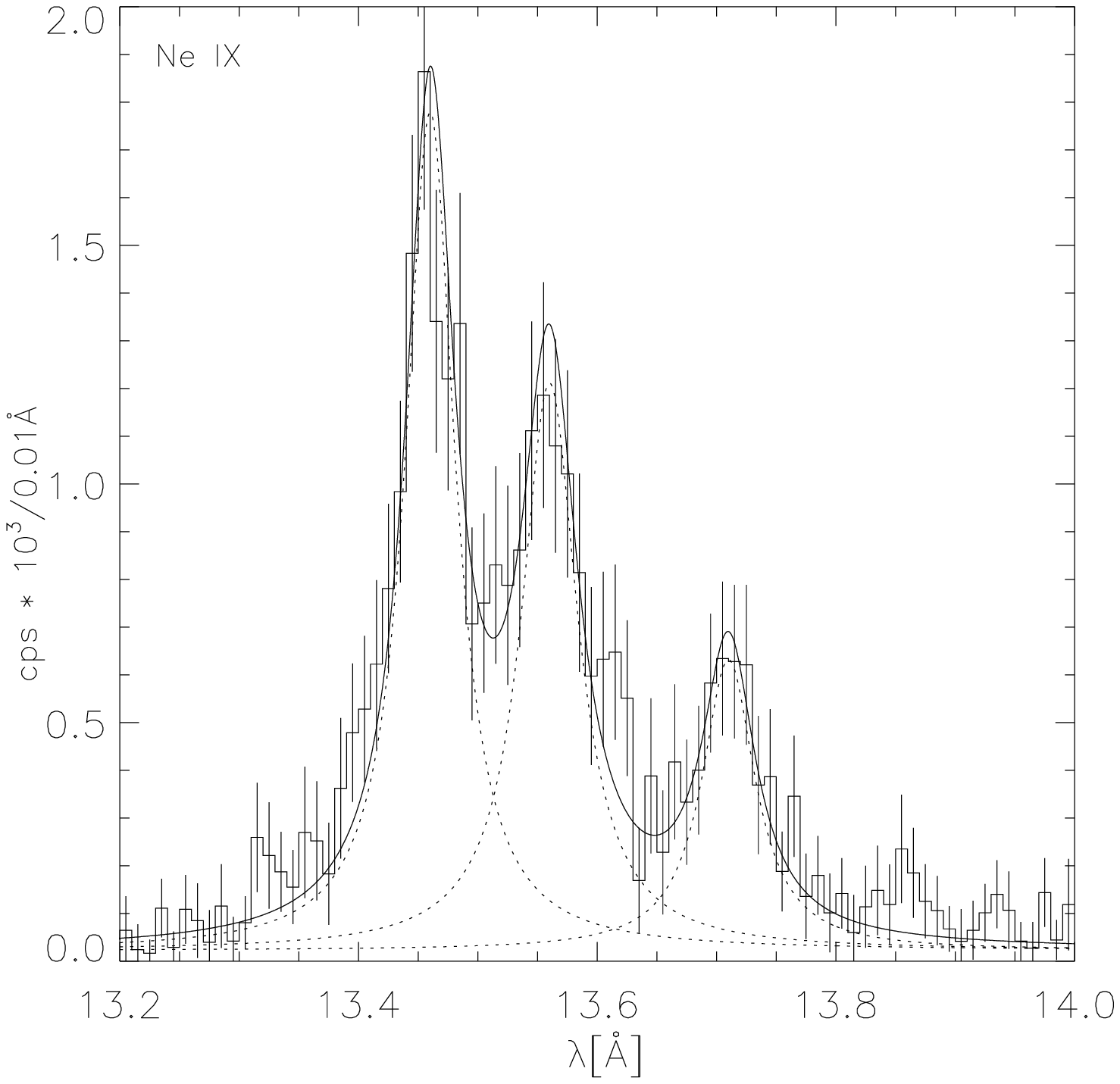}}}
\parbox{9cm}{\resizebox{9cm}{!}{\includegraphics{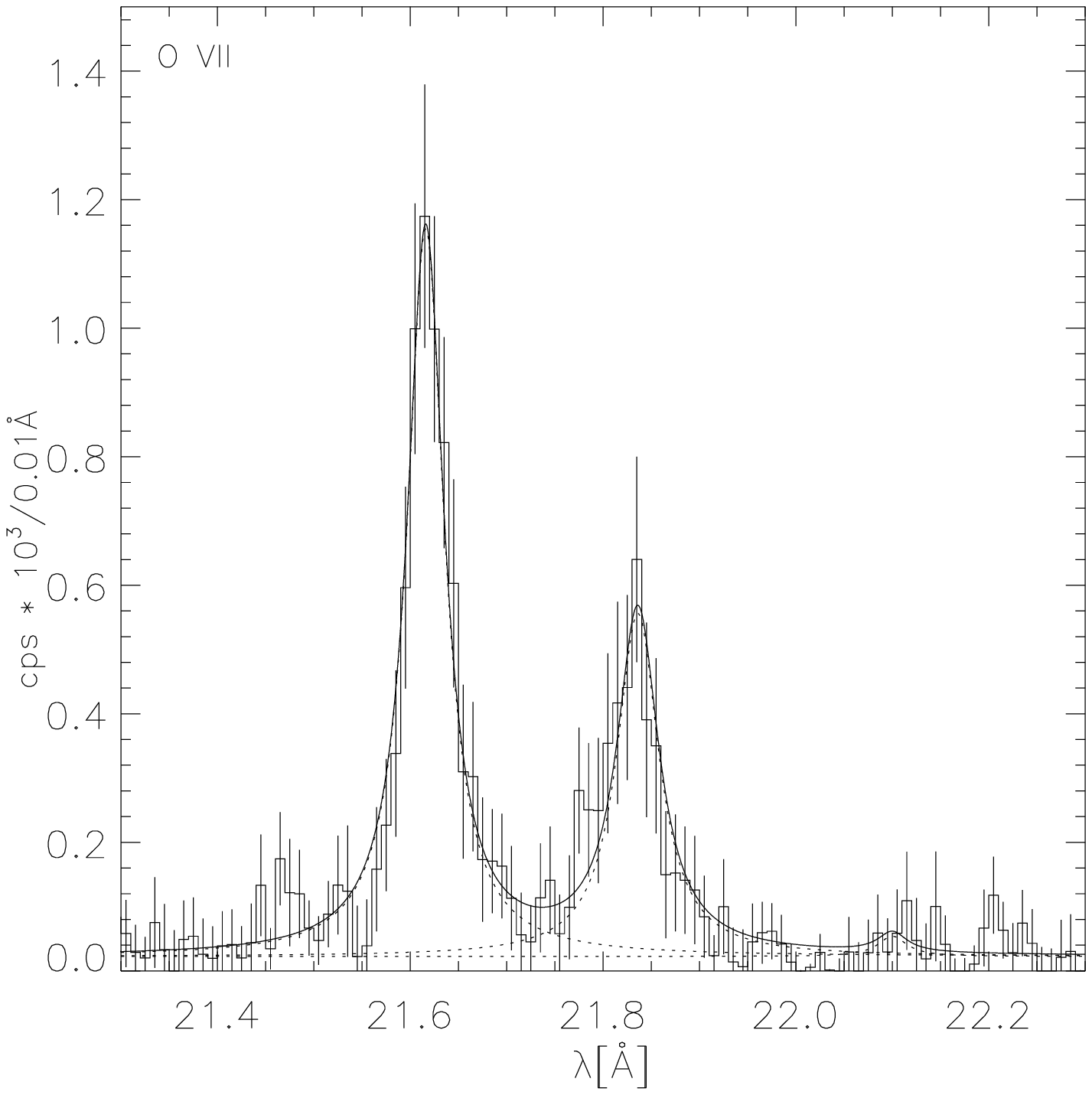}}}
}
\caption{He-like triplets of Ne\,IX and O\,VII
measured with the RGS together with best fit Lorentzians. The continuum
close to each triplet is approximated by a straight horizontal line.
The intensity enhancement to the right of the Ne\,IX triplet in the RGS
spectrum is due to Fe\,XVII. But the major contamination of the Ne\,IX triplet
is expected to come from a Fe\,XIX line which is unresolved from the triplet
intercombination line (see text in Sect.~\ref{subsect:triplets}).}
\label{fig:triplets}
\end{center}
\end{figure*}

The high-resolution X-ray spectrum {obtained with the RGS} allows us
to measure flux ratios
between individual emission lines and use them for plasma
diagnostics. The triplets of He-like ions are of particular importance,
because ratios between the intensity of resonance $r$, intercombination $i$, and
forbidden $f$ line are sensitive to (electron)
temperature (through $G=(f+i)/r$) and density (through $R=f/i$).
While the $G$-ratio is of limited value for temperature diagnostics,
the $R$-ratio constitutes a unique tool
to probe the electron density in stellar coronae.
Although the triplets from silicon (Si\,XIII) to nitrogen (N\,VI)
fall in the spectral range of the RGS,
in TW\,Hya only three elements -- nitrogen, oxygen, and neon -- show sufficient emission in the
respective transitions of their He-like ions to warrant analysis.

In Fig.~\ref{fig:triplets} we zoom in on the spectral region around the oxygen and
neon triplets of TW\,Hya overlaying the data with our line fit result.
For the modeling of both triplets we
used three Lorentzians representing the $r$, $i$, and $f$ lines. Some caution is in order
when interpreting the spectrum near the position of the Ne\,IX triplet: the triplet
is not fully resolved in the RGS, and is known to form a blend with a number of
iron lines, predominantly a Fe\,XIX line at 13.53\,\AA~ that contaminates the Ne
intercombination line.  However, we performed fits of the neon triplet including iron
and find undistinguishable results.  Since we find little hints for the presence of
Fe\,XIX (and generally any iron) in the RGS spectrum
(e.g. the lines at 13.80\,\AA~ and 14.66\,\AA~ are not seen)
we assume that iron contamination plays only a minor role and can be neglected.

The remarkable feature of the He-like triplets of TW\,Hya is the weakness of the
forbidden lines. In the LETGS and RGS spectra of all late-type
stars observed so far the forbidden lines are always stronger than the
intercombination line (in oxygen and neon) and in many cases the measurements
are consistent with the so-called low-density limit
(e.g. \cite{Ness01.1}, \cite{Ness02.1}, \cite{Stelzer02.1}).
This is definitely not the case for TW Hya.
We note in particular that with TW\,Hya being rather cool (spectral type
K7; \cite{Rucinski83.1}) the low intensity of the forbidden line cannot be attributed
to UV radiation at the wavelengths corresponding to
the $f \longrightarrow i$ transitions (\cite{Costa00.1}). Instead it is
a direct indicator of high densities. The formally derived $f/i$-value for {\rm O\,VII}
is $R_{\rm O\,VII} < 0.16$, 
i.e., only an upper limit can be given due to the
weakness of the {\rm O\,VII} $f$-line. Comparing
our measurements to recent calculations by \citey{Porquet01.1} without
any radiation field we find a lower limit on the density of 
$n_{\rm e,O\,VII} > 1\times10^{12}\,{\rm cm^{-3}}$. This is corroborated by the low
flux of the nitrogen forbidden line which is only marginally detected
(cf., Table \ref{tab:rgs_lines}).
The $R$-ratio for neon ($R_{\rm Ne\,IX} = 0.47 \pm 0.11$) 
indicates a density of $n_{\rm e, Ne\,IX} \sim 1\times10^{13}\,{\rm cm^{-3}}$.
Inclusion of the Fe\,XIX blend in the fit of the neon triplet would lower the
strength of the $i$ line, and lead to a somewhat higher value for $R_{\rm Ne\,IX}$
($0.62 \pm 0.12$), and subsequently lower density. 
However the numbers are compatible with the ones cited above
within the statistical uncertainties.

The densities derived from the oxygen and neon
triplets are at least two orders of magnitude above typical coronal densities.
Our density measurements thus fully confirm the results derived by
\citey{Kastner02.1} from their HETGS spectrum, and we conclude in particular that
the high densities cannot be a result of
the flaring reported by \citey{Kastner02.1} in their {\em Chandra} data.

\subsection{Elemental abundances}\label{subsect:abund}

The {\em XMM-Newton} RGS spectrum as well as our spectral fits to the EPIC spectrum
show the X-ray emission of TW\,Hya  clearly dominated by neon emission
in agreement with \citey{Kastner02.1}. Our results specifically suggest a
neon abundance significantly enhanced with respect
to the solar value ($Ne \sim 4\,Ne_\odot$), while iron and oxygen, the
elements with the largest number of lines in the sensitive region of the
{\em XMM-Newton} and {\em Chandra} instruments, are underabundant with respect
to solar values, again in agreement with \citey{Kastner02.1}.
For the elements silicon, magnesium, and nitrogen \citey{Kastner02.1}
claim solar abundances, while carbon lines were not covered by the HETGS spectrum.
Interestingly, the lines of
He-like and H-like silicon are clearly visible in the HETGS spectrum of TW\,Hya
(cf., Fig.~2 in \cite{Kastner02.1}), while the same authors note ``a curious
lack of Mg features, relative to the [i.e. their] model.''  While the RGS spectra
are not sufficiently sensitive in the relevant spectral region, the
EPIC spectrum clearly suggests subsolar abundances also for magnesium and silicon.

\begin{figure*}
\begin{center}
\parbox{18cm}{
\parbox{6cm}{\resizebox{6cm}{!}{\includegraphics{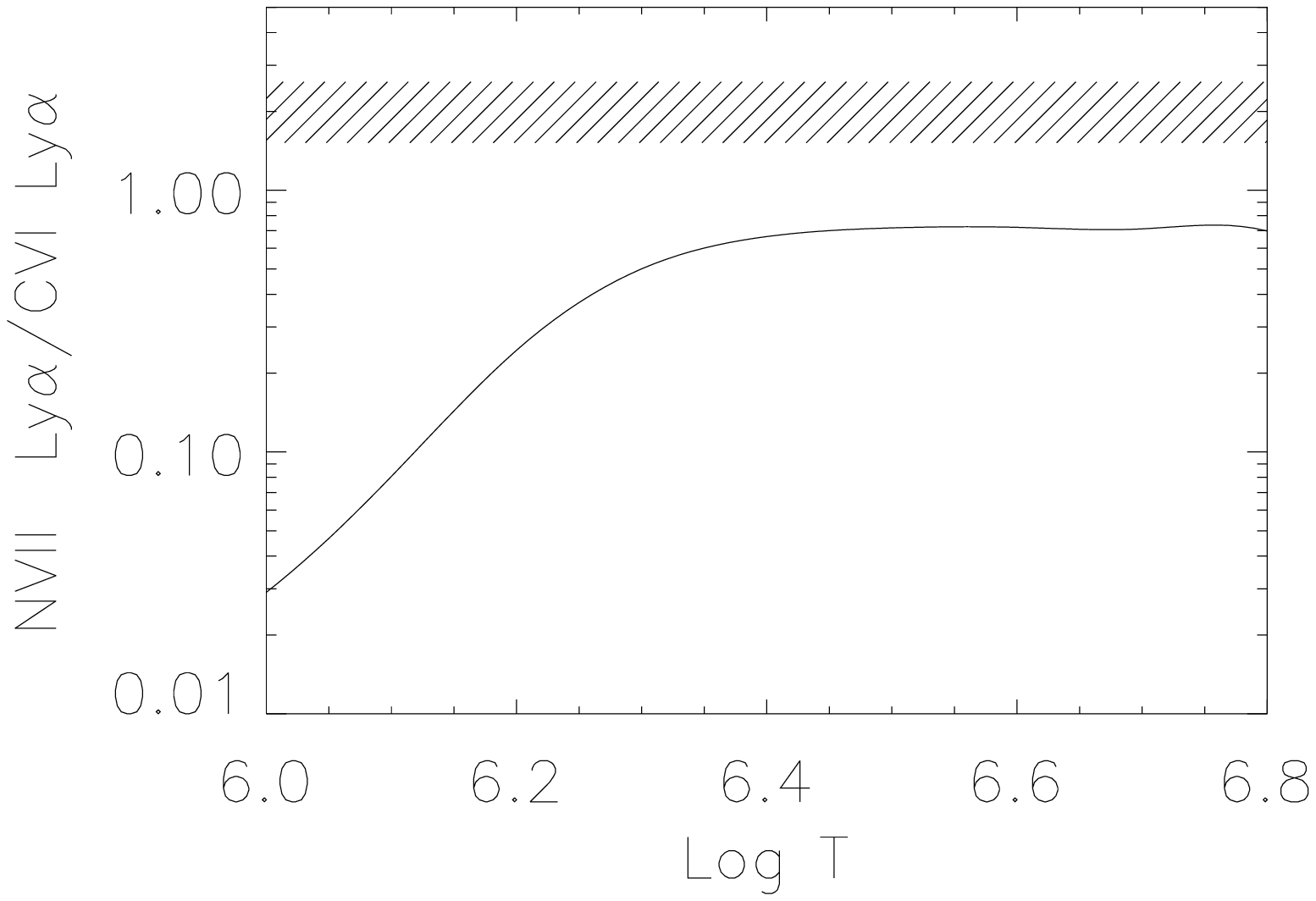}}}
\parbox{6cm}{\resizebox{6cm}{!}{\includegraphics{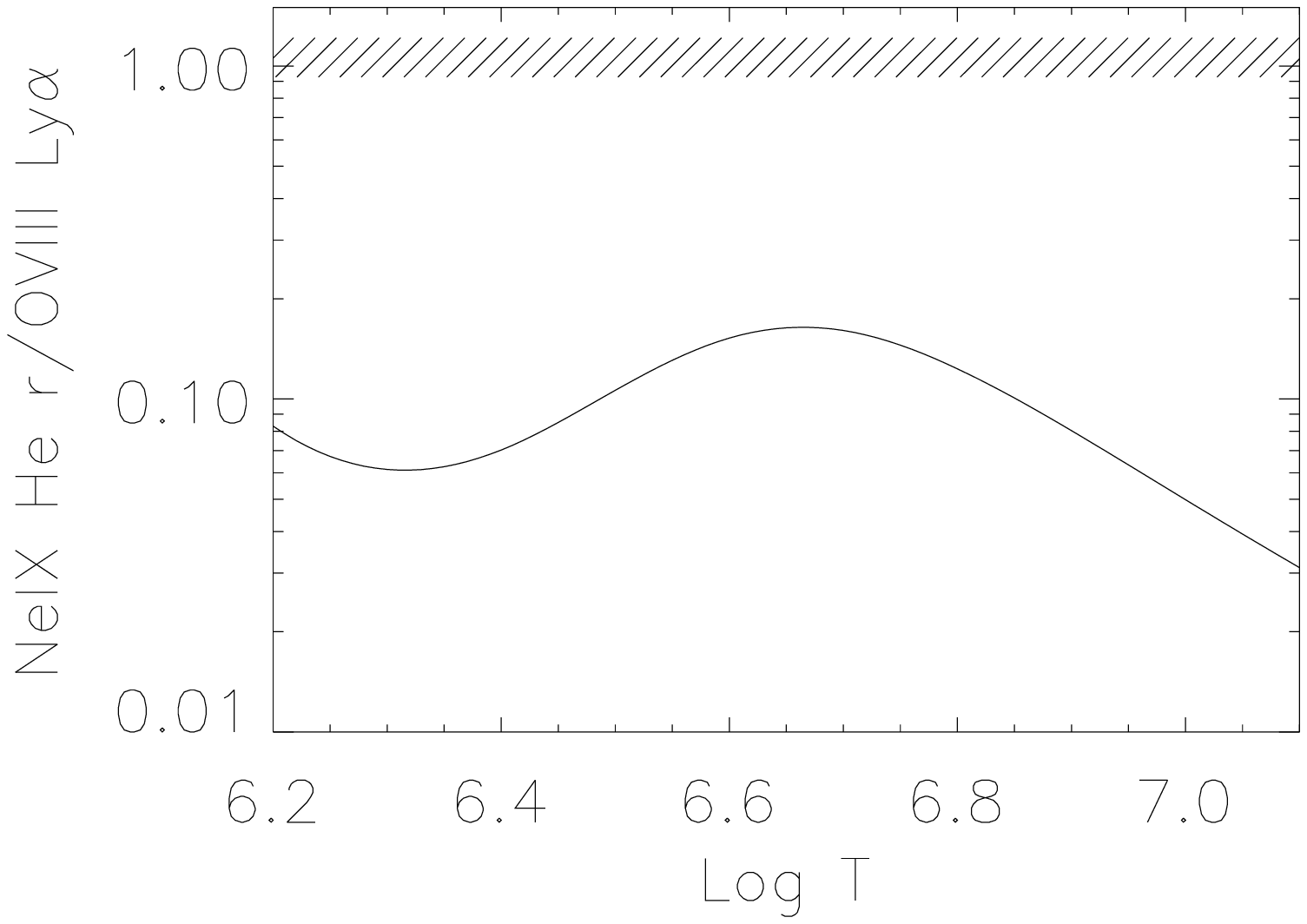}}}
\parbox{6cm}{\resizebox{6cm}{!}{\includegraphics{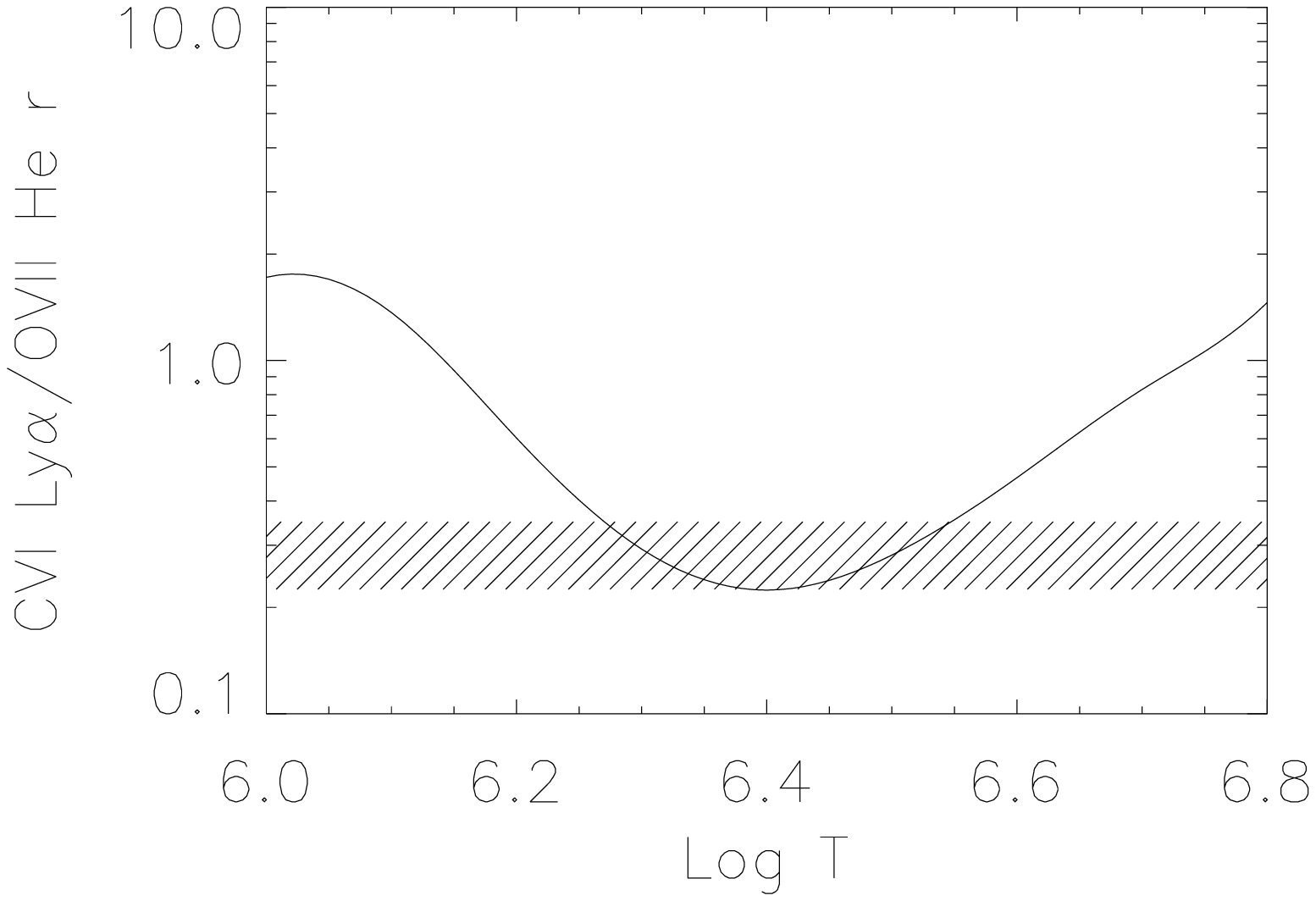}}}
}
\caption{Temperature dependence of various line ratios measured in the RGS spectrum
of TW\,Hya. Solid lines are CHIANTI model calculations (\protect\cite{Dere97.1}), shaded regions
denote the $1\,\sigma$ range observed with the RGS.
(a) - N\,VII $Ly_\alpha$ / C\,VI $Ly_\alpha$;
(b) - Ne\,IX $r$ / O\,VIII $Ly_\alpha$; 
(c) - C\,VI $Ly_\alpha$ / O\,VII $r$.
The plots demonstrate that the C/O abundance ratio is close to solar, while the N/O and Ne/O 
abundance ratio can not possibly be solar.}
\label{fig:temp_ratios2}
\end{center}
\end{figure*}

Let us now focus on the line emission from neon, oxygen, nitrogen, and carbon
as seen by the RGS. Our spectral analysis both with RGS and EPIC shows
that very little low temperature material (with $\log{T}\,{\rm [K]} < 6.3$) can be present.
\citey{Ness02.1} report photon flux ratios for $Ly_\alpha/r$ of 0.59 (oxygen)
and $0.94$ (nitrogen) for the low-temperature F-star Procyon,
and $1.87$ (oxygen) and $3.26$ (nitrogen) for the higher-temperature K-star
$\epsilon$\,Eri, which must be compared to the values of $1.75$ (oxygen) and $4.07$ (nitrogen) 
measured for TW\,Hya despite its overall low temperature.
Although DEM distributions are typically not unique because of the ill-defined nature 
of the inversion problem, we want to emphasize that the peculiar chemical abundances of
TW Hya are not caused by our choice of the EM distribution.
For this we examined the energy flux ratio between the $Ly_\alpha$ lines of
C\,VI and N\,VII as a function of temperature and
find that this line ratio can not be explained by any single temperature or
combination of temperatures under the assumption of solar abundances for both elements
(Fig.~\ref{fig:temp_ratios2}\,a). Therefore nitrogen must be overabundant relative
to carbon (and also oxygen) with respect to the solar abundance ratio, again
in accordance with our fit results.
We then investigated the flux ratio between the $r$ line of neon and the
oxygen $Ly_\alpha$ line in the same way (Fig.~\ref{fig:temp_ratios2}\,b), and again
the observed flux ratio can not be reproduced by any single temperature or
combination of temperatures if the abundances were solar, implying
that neon must be overabundant compared to oxygen with respect to solar abundances,
as found in our fit results. Finally, inspecting the flux ratio between the  
C\,VI $Ly_\alpha$ line and the O\,VII $r$ line we find perfect
consistency assuming the C/O ratio being solar.
Since our EPIC results suggest that silicon, magnesium, iron, and oxygen all
have similar  abundance (with the possible exception of magnesium)
we thus conclude that the elemental abundances of these elements as well
as that of carbon are reduced with respect to solar values by about a factor of $3-4$.
Nitrogen is enhanced with respect to these elements by a factor of $3$ as evidenced
by the {\em XMM-Newton} RGS spectrum,
and neon is enhanced by a factor of $10$ as evidenced by both the
{\em XMM-Newton} RGS and EPIC spectra.
%

\subsection{Interpretation}\label{subsect:inter}

The three properties distinguishing the X-ray spectrum of TW\,Hya from that
of ``normal'' late-type stars are (1) the absence of a hot component
with large emission measure, (2) the high density of the X-ray emitting region,
and (3) the peculiar elemental composition, and in particular the low
abundance of iron. The first two properties can be naturally explained
by attributing the X-ray emission to an accretion shock, the third by the specific
environment of TW\,Hya.

Employing the strong shock formula
$$ T = \frac {3 \mu m_{\rm H} V_{\rm pre}^2} {16 k} $$
with $\mu$ denoting the mean molecular weight, $m_{\rm H}$ the hydrogen atom mass,
and $k$ Boltzmann's constant, we can convert the measured temperature
$\lg{T}\,{\rm [K]} = 6.45$
to an infall preshock velocity $V_{\rm pre}$ of about $350/\sqrt{(\mu)}$\,km/s.
The preshock particle density $n_{\rm pre}$ should be one fourth of the measured post
shock density $n_{\rm post}$, i.e., $2.5 \times 10^{12}\,{\rm cm^{-3}}$.
In order to estimate $L_{\rm post}$, the thickness of
the postshock cooling layer, we equate the infalling kinetic energy flux with
the radiated energy flux (\cite{Lamzin95.1}) through
$$ \frac {\mu m_{\rm H} n_{\rm pre} V_{\rm pre}^3} {2} = n_{\rm post}^2 \Lambda_{\rm mean} L_{\rm post}.$$
Here $\Lambda_{\rm mean}$ denotes the total mean radiative post shock cooling rate
per unit emission measure. Denoting $\Lambda_{\rm mean}$ in units
of 10$^{-23}\,{\rm erg\,cm^3\,s^{-1}}$, we derive
$$ L_{\rm post} = \frac {9 \times 10^{7}} {\sqrt{\mu}} {(\Lambda_{\rm mean,23})}^{-1}\ \ [{\rm cm}].$$
For an isothermal cosmic abundance plasma at $\lg{T}\,{\rm [K]}=6.45$
$\Lambda_{\rm mean,23}$ is generally close to unity;
if we assume  $\Lambda_{\rm mean,23} = 0.3$ because of the depleted elemental abundances
and $\mu = 1.5$ the post-shock thickness is $L_{\rm post} \approx 2500$\,km
with admittedly considerable uncertainty.
Using  the observed O\,VIII energy flux of $3.2 \times 10^{-13}\,{\rm erg/cm^2/s}$ and an
assumed O\,VIII $Ly_{\alpha}$ cooling function of
$9.6 \times 10^{-25}\,{\rm erg\,cm^3\,s^{-1}}$
(assuming $\lg{T}\,{\rm [K]}=6.45$ and an oxygen depletion of $0.3$)
we compute an overall volume emission
measure of $EM = 1.3\times10^{53}\,{\rm cm^{-3}}$ at the distance of $57$\,pc.
Since $n_{\rm post} \approx 10^{13}\,{\rm cm^{-3}}$, the emitting volume $V_{\rm emit}$
must be $V_{\rm emit} = 1.3\times10^{27}\,{\rm cm^{3}}$.
The shock area $A_{\rm shock}$ then becomes
$A_{\rm shock}= V_{\rm emit}/L_{\rm post} = 5.1\times10^{18}\,{\rm cm^2}$,
which would constitute only a very tiny fraction of the visible stellar surface
($< 0.01$\,\%).
Clearly, these estimates are by necessity very rough, but in qualitative
agreement with previous observations in other wavelengths:
\citey{Muzerolle02.1} have shown that the shape of the optical/UV spectral
energy distribution of TW\,Hya can be explained by an accretion shock which
fills $\sim 0.3$\,\% of the stellar surface, and small accretion shocks are also found
on other T Tauri stars (\cite{Calvet98.1}).
Computing the mass accretion rate in the funnel flow through
$M_{\rm acc} = 1/4 A_{\rm shock} n_{\rm post} \mu m_H V_{\rm pre}$
we find $M_{\rm acc} \sim 1 \times 10^{-11}\,{\rm M_\odot/yrs}$,
which is entirely plausible and
within one order of magnitude of the mass accretion rate of TW\,Hya
derived from UV data (\cite{Muzerolle02.1}).

What kind of material is flowing through this funnel onto TW Hya ?
We have shown that this material is
in general metal-depleted in particular with respect to grain-forming
elements such as Mg, Si, Fe, C and O.
Overabundances of neon are found in stellar coronae of RS~CVn systems
(\cite{Audard03.1}), and nitrogen overabundances in evolved stars
(\cite{Schmitt02.1}). The low abundances especially of iron and carbon
call for a different explanation for these chemical peculiarities.
TW\,Hya is rather young and formed ``just recently'' out of a molecular cloud.
Models of cloud chemistry at high densities ($n > 10^7\,{\rm cm^{-3}}$)
predict that almost all metals condense into grains except nitrogen
(\cite{Charnley97.1}).
If the remaining depleted gas is then decoupled from the grains in an accretion disk
around a young star (as is expected in planetary formation models),
the accreted and shock-heated gas
is expected to be metal-deficient as well.
Indeed, \citey{Weinberger02.1} found evidence for the presence for condensed
silicates in the disk of TW\,Hya, and
\citey{Herczeg02.1} speculated about a deficiency of Si in the accretion
flow of TW\,Hya  based on the examination of its far-UV spectrum.
{\em XMM-Newton} has completed this picture of TW\,Hya as an
accreting PMS star, and for the first time provided a consistent view
of X-ray emission from a stellar accretion shock by observation.

\section{Summary}\label{sect:summary}

To summarize, the marked differences between the X-ray properties of TW\,Hya
and active main-sequence dwarf stars suggest either
(i) strong evolutionary effects on coronal structures, or (ii) an entirely
different origin for the X-ray emission of TW\,Hya.
X-ray emission from a shock at the bottom of an accretion
column provides a plausible mechanism for the latter possibility. The derived shock
temperatures, gas densities, filling factors and mass accretion rates as well as
the obvious lack of grain-forming elements in the X-ray emitting region
provide strong indications in favor of this scenario.

A high-resolution X-ray spectrum of another member of the TW\,Hya
association, HD\,98800, was recently obtained with {\em Chandra} 
(CXC Press Release 03-04).
The object is a quadruple star with a circumbinary disk. In terms of
H$_\alpha$ emission HD\,98800 is a wTTS. 
Therefore it does not come as a surprise that its X-ray properties
are very different from those of TW\,Hya, and
clearly reminiscent for typical stellar activity. 
Similarly, the high-resolution X-ray observation of PZ\,Tel,
a $\sim 20$\,Myr old post-TTS in the Tucanae association, does not show properties
drammatically different from older late-type stars (Argiroffi et~al., in prep.).

We conclude that TW\,Hya stands out as a presently unique
example for a star with strong evidence for accretion signatures
in its X-ray spectrum. 
Future high spectral resolution and high sensitivity
X-ray observations will have to show whether soft emission,
high densities, and metal-deficiency are general characteristics of cTTS.

\begin{acknowledgements}

BS acknowledges financial support from the European Union by the Marie
Curie Fellowship Contract No. HPMD-CT-2000-00013. X-ray astronomy at
the Hamburger Sternwarte is generously supported by DLR.
We have used the CORA line fitting package kindly provided by J.-U. Ness.
This work is based on observations obtained with {\em XMM}-Newton,
an ESA science mission with instruments and contributions directly
funded by ESA Member States and the USA (NASA).

\end{acknowledgements}

\end{document}